\documentclass{elsart}
\journal{Physics Letter A}

\usepackage{graphicx,color}
\graphicspath{{figures/}}

\usepackage{amssymb,amsmath,bm}
\usepackage{cite}
\usepackage{url}

\newtheorem{definition}{Definition}
\newtheorem{property}{Property}

\newtheorem{theorem}{Theorem}
\newtheorem{remark}{Remark}
\newenvironment{proof}{\noindent\textit{Proof}: }{\hfill$\blacksquare$\vskip 0.5\baselineskip}
\newcommand\mymatrix[1]{\bm{\mathrm{#1}}}

\begin{document}

\begin{frontmatter}

\title{Cryptanalysis of two chaotic encryption schemes based on circular bit shift and XOR
operations}
\thanks{This paper has been accepted by \textit{Physics
Letters A} in March 2007, DOI: 10.1016/j.physleta.2007.04.023.}

\author[hk-cityu]{Chengqing Li\corauthref{corr}},
\author[germany]{Shujun Li\corauthref{corr}},
\author[spain]{Gonzalo Alvarez},
\author[hk-cityu]{Guanrong Chen} and
\author[hk-polyu]{Kwok-Tung Lo}

\corauth[corr]{Corresponding authors: Chengqing Li
(swiftsheep@hotmail.com), Shujun Li (http://www.hooklee.com).}

\address[hk-cityu]{Department of Electronic Engineering, City University of Hong Kong,
83 Tat Chee Avenue, Kowloon Tong, Hong Kong SAR, China}
\address[germany]{FernUniversit\"{a}t in Hagen, Lehrgebiet Informationstechnik, Universit\"{a}tsstra{\ss}e 27, 58084 Hagen, Germany}
\address[spain]{Instituto de F\'{\i}sica Aplicada, Consejo Superior de
Investigaciones Cient\'{\i}ficas, Serrano 144, 28006 Madrid, Spain}
\address[hk-polyu]{Department of Electronic and Information Engineering,
The Hong Kong Polytechnic University, Hung Hom, Kowloon, Hong Kong
SAR, China}

\begin{abstract}
Recently two encryption schemes were proposed by combining
circular bit shift and XOR operations, under the control of a
pseudorandom bit sequence (PRBS) generated from a chaotic system.
This paper studies the security of these two encryption schemes
and reports the following findings: 1) there exist some security
defects in both schemes; 2) the underlying chaotic PRBS can be
reconstructed as an equivalent key by using only two chosen
plaintexts; 3) most elements in the underlying chaotic PRBS can be
obtained by a differential known-plaintext attack using only two
known plaintexts. Experimental results are given to demonstrate
the feasibility of the proposed attack.
\end{abstract}

\end{frontmatter}

\section{Introduction}

In the past three decades, many digital chaotic ciphers
\cite{Kocarev:ISCAS98, Alvarez:ICST99,
Silva:Chaos-Communications:AeroSpace2000,
Kocarev:ChaosCryptography:IEEECASM2001,
Schmitz:ChaoticCryptography:JFI2001, ShujunLi:Dissertation2003,
Li:ChaosImageVideoEncryption:Handbook2004, Lcq:MasterThesis2005,
LiShujun:TalksCC2:BJNU} and analog chaos-based secure
communication schemes \cite{Hasler:Survey:IJBC98, Alvarez:ICST99,
Silva:Chaos-Communications:AeroSpace2000, Yang:Survey:IJCC2004,
LiShujun:TalksCC1:SZU} have been proposed, trying to explore the
intrinsic relationship between chaos and cryptography. However,
due to the lack of a strict scrutiny on the security, most
chaos-based cryptosystems have been found insecure against various
attacks \cite{ShujunLi:Dissertation2003,
Li:ChaosImageVideoEncryption:Handbook2004, LiShujun:TalksCC2:BJNU,
LiShujun:TalksCC1:SZU, Lcq:MasterThesis2005,
LiShujun:Rules:IJBC2006}.

In \cite{Taoxiang:PLA2006}, a new block encryption scheme was
proposed by combining circular bit shift and XOR operations, under
the control of a pseudorandom bit sequence (PRBS) generated from
the chaotic logistic map. Later, in \cite{Wenwu:PLA2006}, the
above encryption scheme was further modified\footnote{Note that
the authors of \cite{Wenwu:PLA2006} did not make clear that their
work is a modification of the one proposed in
\cite{Taoxiang:PLA2006}. However, it is obvious that the
DCNN-based scheme in \cite{Wenwu:PLA2006} was originated from the
work reported in \cite{Taoxiang:PLA2006} because the encryption
procedures of the two schemes are exactly the same except for some
minor modifications.}, by adopting some alterations such as
replacing the logistic map with a delayed chaotic neural network
(DCNN).

In \cite[Sec.~6.2]{Taoxiang:PLA2006}, it is pointed out that the
encryption scheme based on the logistic map is not secure enough
against chosen-plaintext attack, because there exists some
information leakage about the chaotic trajectory involved. Then,
the authors of \cite{Taoxiang:PLA2006} suggested using key
switching and/or ``cycling chaos"
\cite{Palacios:CyclingChaosCryptography:PLA2002} as remedies to
further improve the security. However, as we show below in this
paper, the information leakage is actually not the main reason why
the encryption scheme is not secure against chosen-plaintext
attack. We further show that both encryption schemes are not only
insecure against chosen-plaintext attack, but also insecure
against a differential known-plaintext attack. In addition, we
will point out some other security defects existing in the design
of these two chaos-based encryption schemes.

The rest of the paper is organized as follows. The next section
gives a brief introduction to the two encryption schemes. Some
security problems existing in both of the two encryption schemes
are reported in Sec.~\ref{sec:securityproblems}. The main
cryptanalytic results about plaintext attacks are given in
Sec.~\ref{sec:Cryptanalysis}, with some experimental results for
demonstration. The last section concludes this paper.

\section{Two Chaotic Encryption Schemes}
\label{sec:scheme}

To facilitate the following description of the two encryption
schemes, the definitions of circular bit shift operations and some
notations are first introduced.

\begin{definition}\label{definition:lll_ggg}
Assuming that $L\in\mathbb{Z}^+$, $x\in\mathbb{Z}$ and
$a=\sum_{i=0}^{L-1}\left(a_i\cdot 2^i\right)\in\{0,\cdots,2^L-1\}$,
where $a_i\in\{0,1\}$, the $L$-bit left and right circular bit shift
operations are defined as follows: $a\lll_L
x=a\ggg_L(-x)=\sum_{i=0}^{L-1}\left(a_i\cdot 2^{(i+x)\bmod
L}\right)$ and $a\ggg_L x=a\lll_L(-x)=\sum_{i=0}^{L-1}\left(a_i\cdot
2^{(i-x)\bmod L}\right)$.
\end{definition}

From Definition~\ref{definition:lll_ggg}, one can easily verify
some simple properties about the circular bit shift operations: 1)
$\forall\ x\equiv 0\pmod L$, $a\lll_L x=a\ggg_L x=a$; 2) $\forall
x_1\equiv x_2\pmod L$, $a\lll_L x_1=a\lll_L x_2$ and $a\ggg_L
x_1=a\ggg_L x_2$; 3) $\forall x_1\equiv x_2\pmod L$, $(a\lll_L
x_1)\ggg_L x_2=(a\ggg_L x_1)\lll_L x_2=a$. The proofs are simple,
therefore omitted. These properties will be directly used
hereinafter without further explanations.

Both encryption schemes work with $L$-bit blocks (some zero bits
are padded when the last plain-block contains less than $L$ bits).
In \cite{Taoxiang:PLA2006} $L=64$ and in \cite{Wenwu:PLA2006}
$L=32$, so the two encryption schemes are 64-bit and 32-bit block
ciphers, respectively. Throughout the paper, we assume that the
plaintext contains $N$ blocks: $\{P_j\}_{j=0}^{N-1}$, and the
corresponding ciphertext is $\{C_j\}_{j=0}^{N-1}$.

\subsection{Encryption Scheme Based on the Logistic Map \cite{Taoxiang:PLA2006}}

In this scheme, \textit{the secret key} is the initial condition
$x(0)$ and control parameter $\mu$ of the following chaotic
logistic map:
\begin{equation}\label{eq:Logistic}
f(x)=\mu x(1-x).
\end{equation}

The core of the encryption scheme is a PRBS,
$\{B_i\}_{i=0}^{70N-1}$, which is generated from the chaotic
logistic map. Two pseudorandom number sequences (PRNS),
$\{A_j\}_{j=0}^{N-1}$ and $\{D_j\}_{j=0}^{N-1}$, are further
derived from the PRBS for the encryption/decryption purpose. The
whole procedure can be described in the following
steps\footnote{To give a clearer and simpler description, we
change some notations used in \cite{Taoxiang:PLA2006}.}.

\begin{itemize}
\item \textit{Step 1}: Set $j=0$, $r=3$ and iterate the logistic
map from $x(0)$ for $N_0=250$ times.

\item \textit{Step 2}: Iterate the logistic map for $N_1=70$ times
to get a sequence composing of 70 chaotic states. Then, extract
the $r$-th bit from each chaotic state's binary representation to
get 70 pseudorandom bits $\{B_i\}_{i=70j}^{70j+69}$.

\item \textit{Step 3}: Set
$A_j=\sum_{k=0}^{63}\left(B_{70j+k}\cdot 2^{63-k}\right)$,
$D_j=\sum_{k=64}^{69}\left(B_{70j+k}\cdot 2^{69-k}\right)$ and
$j=j+1$.

\item \textit{Step 4}: If $j\leq N-1$, iterate the logistic map
for $D_j$ times and then goto \textit{Step 2}; otherwise, stop the
process.
\end{itemize}

After the two PRNS $\{A_j\}_{j=0}^{N-1}$ and $\{D_j\}_{j=0}^{N-1}$
have been determined, \textit{the encryption procedure} can be
described easily by the following equation:
\begin{equation}\label{equation:encryption1}
C_j=(P_j\lll_{64} D_j)\oplus A_j.
\end{equation}
Accordingly, the decryption procedure is as follows:
\begin{equation}\label{equation:decryption1}
P_j=(C_j\oplus A_j)\ggg_{64} D_j.
\end{equation}

\subsection{Encryption Scheme Based on a Delayed Chaotic Neural Network
\cite{Wenwu:PLA2006}}

Compared with the scheme introduced in the last subsection, the
DCNN-based one can be described as follows.

\begin{enumerate}
\item The chaotic system is replaced by a DCNN with $n=2$ neurons,
described by the following equation:
\begin{equation}\label{eq:twoneurons}
{\dot{x_1}(t)\choose\dot{x_2}(t)} =-\mymatrix{C}{x_1(t) \choose
x_2(t)}+\mymatrix{A}{\tanh(x_1(t))\choose \tanh(x_2(t))}
+\mymatrix{B}{\tanh(x_1(t-\tau(t))\choose \tanh(x_2(t-\tau(t))},
\end{equation}
where $(x_1(t),x_2(t))^{T}\in \mathbb{R}^2$ is the state vector
associated with the 2 neurons, $\tau(t)$ is a time-delay function,
$\mymatrix{C}=\mbox{diag}(c_1, c_2)$ is a diagonal matrix,
$\mymatrix{A}=[a_{i,j}]_{2\times 2}$,
$\mymatrix{B}=[b_{i,j}]_{2\times 2}$ are the connection weight
matrix and the delayed weight matrix, respectively.

As the main cryptanalysis given in this paper does not depend on
this chaotic neural network, more details about this
$n$-dimensional chaotic system are referred to
\cite[Sec.~2]{Wenwu:PLA2006}. Since the chaotic neural network is
an analogue dynamical system, it has to be approximated by a
discrete-time one by using a numerical algorithm with time step
$h$.

\item
The secret key was claimed to include the initial condition and
control parameters of the DCNN, the value of $h$, the structure of
the DCNN and the numerical algorithm that implements the DCNN.

\item One neuron of the DCNN is selected to generate a shorter
PRBS, $\{B_i\}_{i=0}^{38N-1}$, for the encryption of each
plain-block. The generation process of the PRBS is now changed as
follows, where $s$ is used to choose one neuron for encryption of
the next plain-block.
\begin{itemize}
\item \textit{Step 1}: Set $j=0$, $r=4$, $s=1$, and iterate the
DCNN from its initial condition for $N_0=1000$ time steps.

\item \textit{Step 2}: Iterate the DCNN for $N_1=38$ time steps.
For each state of the $s$-th neuron, scale it to be within the
unit interval $[0,1]$ and then extract the $r$-th bit from the
binary representation of the scaled state, so as to get 38
pseudorandom bits $\{B_i\}_{i=38j}^{38j+37}$.

\item \textit{Step 3}: Set
$A_j=\sum_{k=0}^{31}\left(B_{38j+k}\cdot 2^{31-k}\right)$,
$D_j=\sum_{k=32}^{36}\left(B_{38j+k}\cdot 2^{36-k}\right)$,
$s=B_{38j+37}+1$ and $j=j+1$.

\item \textit{Step 4}: If $j\leq N-1$, iterate the DCNN for $D_j$
time steps and then goto \textit{Step 2}; otherwise, stop the
process.
\end{itemize}

\item
An extra bit shift operation is introduced on $A_j$. By doing that,
the encryption procedure becomes
\begin{equation}\label{equation:encryption2}
C_j=(P_j\lll_{32} D_j)\oplus(A_j\ggg_{32} D_j).
\end{equation}
Similarly, the decryption procedure is changed to
\begin{equation}\label{equation:decryption2}
P_j=(C_j\oplus(A_j\ggg_{32} D_j))\ggg_{32} D_j.
\end{equation}
\end{enumerate}

\section{Some Security Problems}
\label{sec:securityproblems}

\newlength\figwidth
\setlength\figwidth{0.32\columnwidth}

\subsection{Insufficient Randomness of Chaos-Based PRBS $\{B_i\}$}
\label{ssec:nonuniformdistribution}

In both encryption schemes, it is expected that the PRBS is random
enough to ensure a high level of security. However, as shown
below, neither the chaotic trajectories of the logistic map nor
those of the DCNN have a uniform distribution, which leads to
insufficient randomness of the PRBS generated from these chaotic
trajectories.

\begin{figure}[!htb]
\center
\begin{minipage}{2\figwidth}
\includegraphics[width=\textwidth]{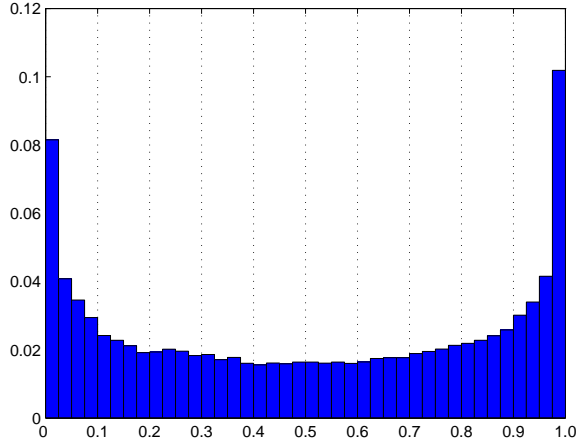}
\end{minipage}\\
\caption{A typical distribution of some random chaotic
trajectories of the logistic map with control parameter
$\mu=3.999$.} \label{figure:logisticdistribution}
\end{figure}

For the logistic map, distributions of a number of chaotic
trajectories, generated by iterating Eq.~(\ref{eq:Logistic}) for
$10^5$ times with random initial conditions and random control
parameters, were studied. All the distributions are quite close to
each other, so only one typical example is shown in
Fig.~\ref{figure:logisticdistribution} for illustration.
Apparently, the non-uniform distribution of the chaotic trajectory
will inevitably degrade the randomness of the derived PRBS
$\{B_i\}$. For verification, we employed the NIST statistical test
suite \cite{Rukhin:TestPRNG:NIST} to test the randomness of 100
binary sequences of length $\frac{256\cdot 256}{8}\cdot 70=573440$
(the number of bits used for encryption of a $256\times 256$ plain
gray-scale image). Note that the 100 binary sequences were
generated with randomly selected secret keys. For each test, the
default significance level 0.01 was used. The results are shown in
Table~\ref{table:test}, from which one can see that the PRBS
$\{B_i\}$ does not satisfy the requirements as a good random
source.

\begin{table}[!htbp]
\center \caption{The performed tests and the number of sequences
passing each test in a sample of 100 sequences.}\label{table:test}
\begin{tabular}{c|c}
\hline Name of Test & Number of Passed Sequences\\
\hline\hline Frequency & 0\\
\hline Block Frequency & 3\\
\hline  Cumulative Sums & 0\\
\hline Runs & 0\\
\hline Rank  & 82\\
\hline Discrete Fourier Transform & 32\\
\hline Non-overlapping Template Matching& 0\\
\hline Serial & 0\\
\hline Approximate Entropy& 0\\
\hline
\end{tabular}
\end{table}

\begin{figure}[!htb]
\center
\begin{minipage}{2\figwidth}
\includegraphics[width=\textwidth]{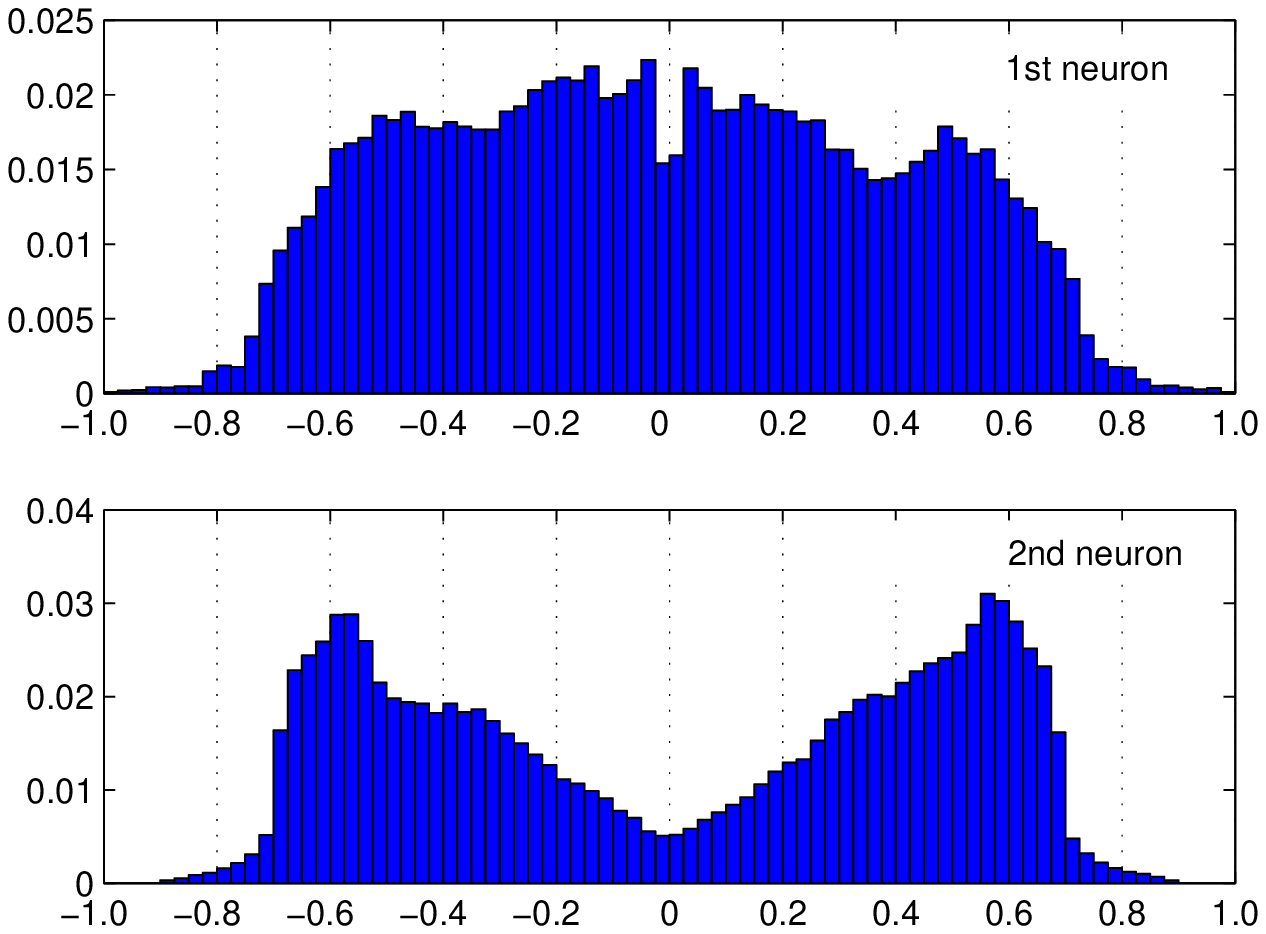}
\end{minipage}\\
\caption{The empirical distributions of the two neurons' states of
the DCNN Eq.~(\ref{eq:twoneurons}).}
\label{figure:dcnndistribution}
\end{figure}

For the DCNN used in \cite{Wenwu:PLA2006}, the non-uniformity of
the chaotic trajectory corresponding to each neuron is even worse.
Observing Fig.~1 in \cite{Wenwu:PLA2006}, one can easily see that
the trajectory seldom visits some regions of the phase space. We
performed some experiments to further investigate this
problem\footnote{In the experiments, the involved delay
differential equation (DDE) with a time-varying delay was
numerically solved by the method proposed in
\cite{Shampine:DDESDCODE} with the same default error tolerance.}.
Figure~\ref{figure:dcnndistribution} shows the distributions of
622,600 chaotic states of the two neurons of the DCNN with the
following configurations: the initial condition $\bm{x}(t\leq
0)=(0.4,0.6)^T$, $\tau(t)=1+0.1\sin(t)$, and the three matrices in
Eq.~(\ref{eq:twoneurons}) were set as follows:
\[
\mymatrix{A}=\left(\begin{matrix}%
2 & -0.1\\
-5 & 3
\end{matrix}\right),
\mymatrix{B}=\left(\begin{matrix}%
-1.5 & -0.1\\
-0.2 & -2.5
\end{matrix}\right),
\mymatrix{C}=\left(\begin{matrix}%
1 & 0\\
0 & 1
\end{matrix}\right).
\]
The time step size $h=0.01$ was used in the numerical solution to
simulate the DCNN.

The distributions shown in Fig.~\ref{figure:dcnndistribution}
imply that the randomness of the PRBS $\{B_i\}$ is weaker than
that derived from the logistic map. Moreover, there exists another
more serious problem that dramatically influences the randomness
of the PRBS derived from the DCNN. As can be seen, the DCNN is an
analogue dynamical system with a continuous trajectory, which
means that any two consecutive chaotic states simulated via a
numerical algorithm are always closely correlated. As a result,
the chaotic bits derived from consecutive chaotic states will also
be closely correlated. Furthermore, the smaller the time step size
$h$ is, the stronger such a correlation will be. However, as
mentioned in the last subsection, the time step size $h$ should be
small enough to achieve a good estimation of the true dynamics of
the DCNN. That is, the close correlation between consecutive bits
is an unavoidable defect of PRBS based on any analogue dynamical
system like this DCNN.

To evaluate the real randomness of the PRBS $\{B_i\}$ derived from
the DCNN, we carried out the runs test
{\cite[Sec.~5.4.4]{MOV:CyrptographyHandbook1996} on the first
20,000 bits of $\{B_i\}$ corresponding to the trajectory shown in
Fig.~\ref{figure:dcnndistribution}, where the definition of run of
a binary sequence is given in
\cite[Sec.~2.3.1]{Rukhin:TestPRNG:NIST}: ``\textit{A run of length
k consists of exactly k identical bits and is bounded before and
after with a bit of the opposite value}". The result is shown in
Fig.~\ref{figure:runstest}. As a comparison, the mathematical
expectations of the number of runs of various lengths in an ideal
random binary sequence are also plotted. Observing
Fig.~\ref{figure:runstest}, one can see that the randomness of the
PRBS generated by the DCNN is obviously very weak. As a result of
the poor randomness of $\{B_i\}$, it is expected that $\{A_j\}$
and $\{D_j\}$ are also far from being random, which can be clearly
seen by looking at the numbers of different values in
$\left\{\left\lfloor
A_j/2^{22}\right\rfloor\right\}_{j=0}^{16383}$ and
$\{D_j\}_{j=0}^{16383}$, as shown in
Fig.~\ref{figure:distributevariable}.

\begin{figure}[!htb] \center
\begin{minipage}{2\figwidth}
\includegraphics[width=\textwidth]{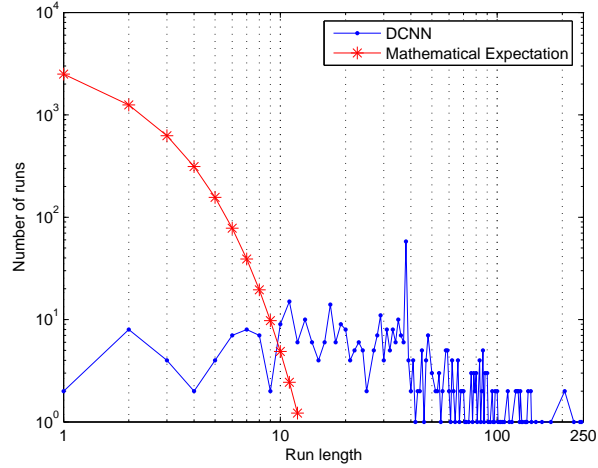}
\end{minipage}
\caption{The numbers of $k$-bit runs in one PRBS
$\{B_i\}_{i=0}^{19999}$ generated by the DCNN versus the expected
numbers of a random bit sequence, where $k=1\sim
250$.}\label{figure:runstest}
\end{figure}

\begin{figure}[!htb]
\center
\begin{minipage}{2\figwidth}\centering
\includegraphics[width=\textwidth]{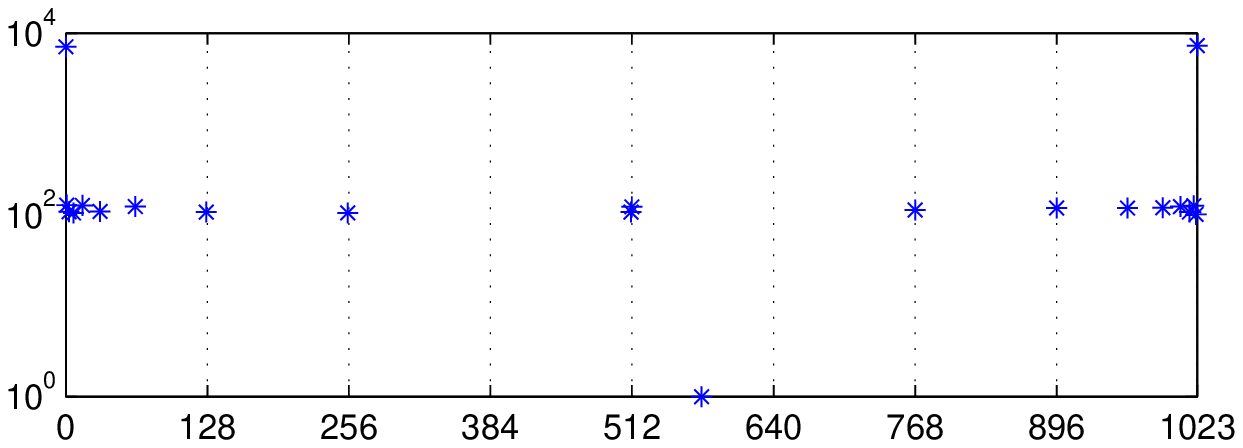}
a)
\end{minipage}
\\
\begin{minipage}{2\figwidth}\centering
\includegraphics[width=\textwidth]{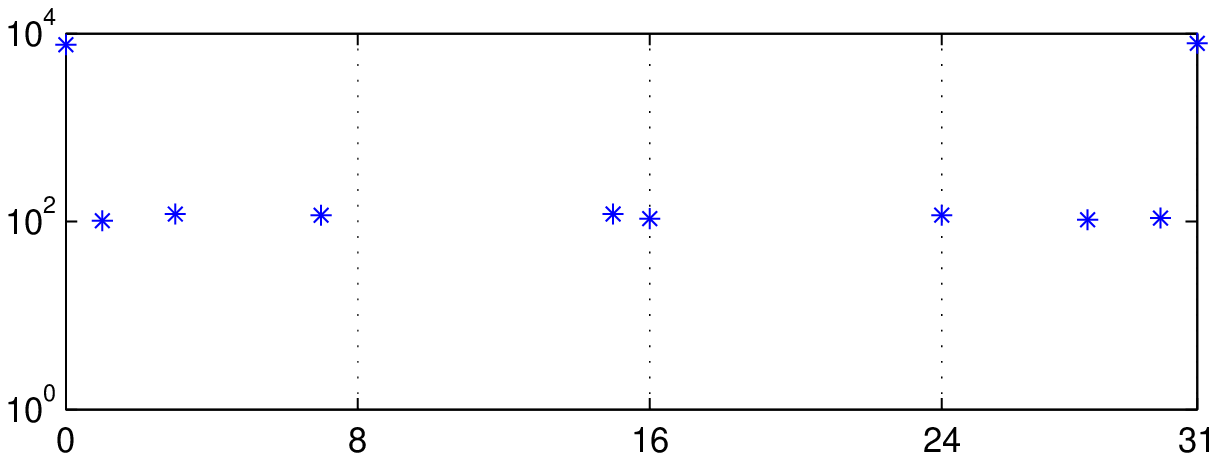}
b)
\end{minipage}
\caption{The numbers of different values of a)
$\left\{\left\lfloor
A_j/2^{22}\right\rfloor\right\}_{j=0}^{16383}$ and b)
$\{D_j\}_{j=0}^{16383}$. Note that only the numbers of values
existing in the sequences are plotted in the two
sub-figures.}\label{figure:distributevariable}
\end{figure}

Due to the serious non-uniform distributions of $\{A_j\}$ and
$\{D_j\}$ in the DCNN-based scheme, it is suspected that the
encryption performance may not be satisfactory. For example,
Fig.~\ref{figure:distributevariable} shows that the probability
that $A_j=D_j=0$ is relatively high, which means that the
encryption totally fails in this case. For two typical images,
``Lenna" and ``Peppers", the encryption results of the DCNN-based
scheme are shown in Fig.~\ref{figure:cipherimages-DCNN}, where the
secret key was set to be the one used for
Fig.~\ref{figure:dcnndistribution}. One can see that some visual
information about the plain-images has been leaked from the
cipher-images. As a comparison, the encryption results of the
scheme based on the logistic map is given in
Fig.~\ref{figure:cipherimages-logistic}, where the original secret
key in \cite{Taoxiang:PLA2006} was used: $x(0)=0.1777$,
$\mu=3.9999995$. Comparing Figs.~\ref{figure:cipherimages-DCNN}
and \ref{figure:cipherimages-logistic}, one can see that the
DCNN-based scheme has a much worse encryption performance than the
one based on the logistic map.

\begin{figure}[!htb]
\centering
\begin{minipage}[t]{\figwidth}
\centering
\includegraphics[width=\figwidth]{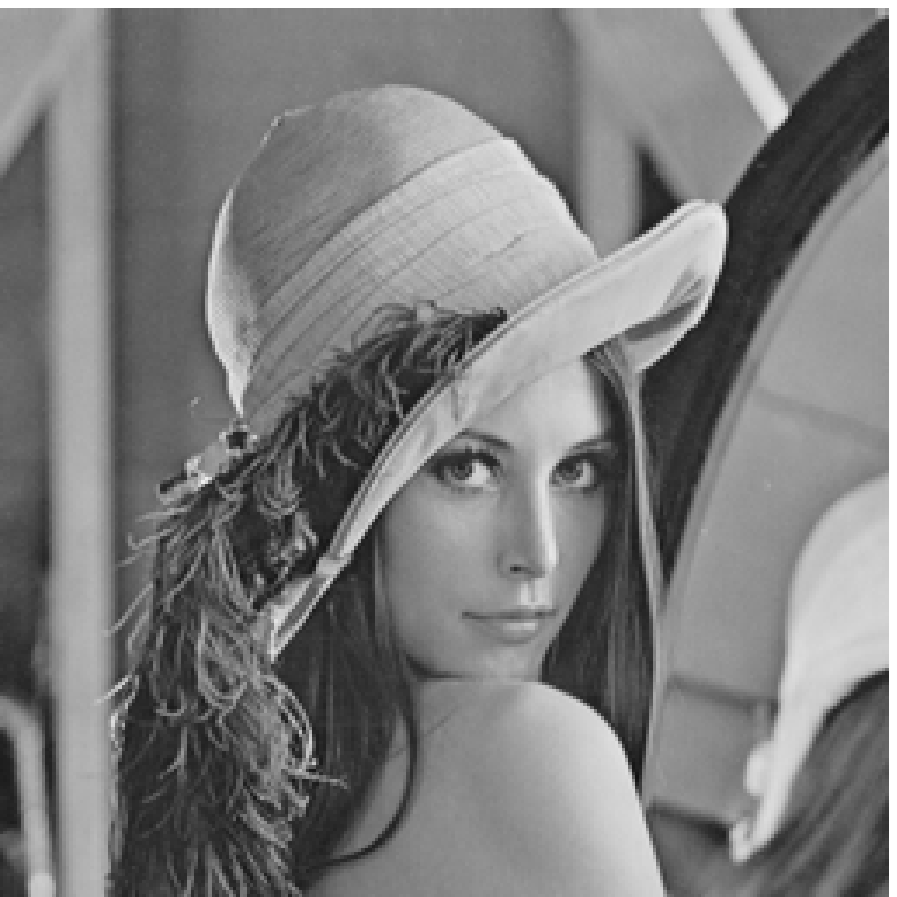}
a)
\end{minipage}
\begin{minipage}[t]{\figwidth}
\centering
\includegraphics[width=\figwidth]{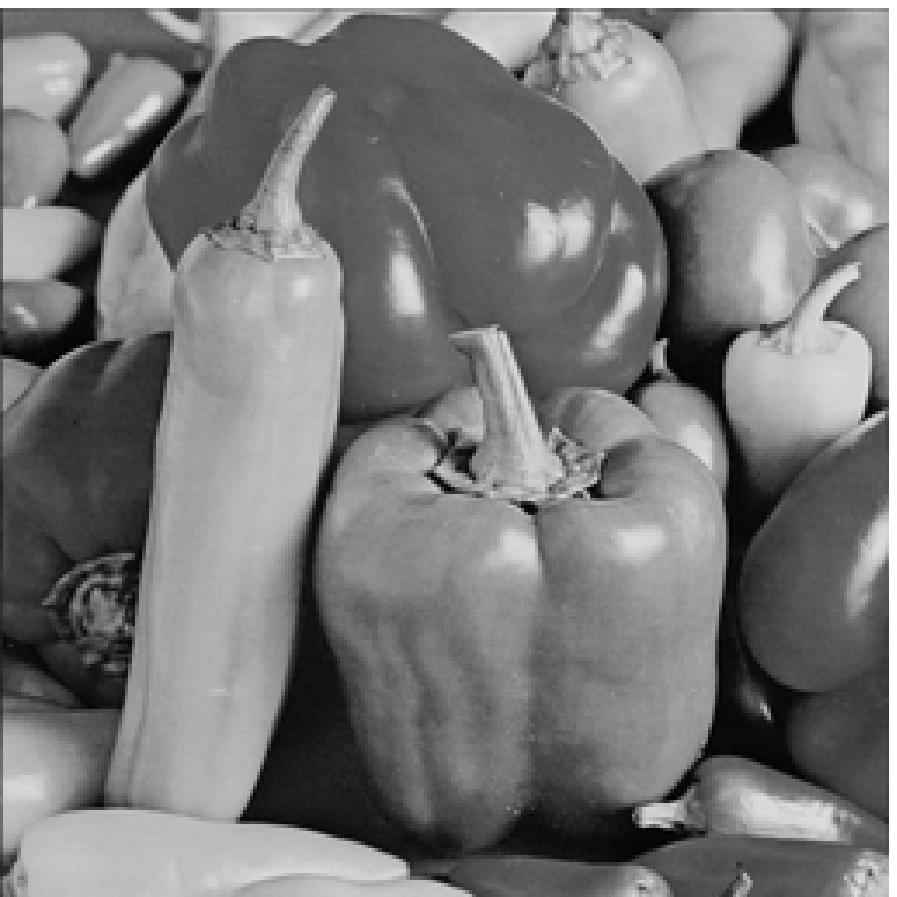}
b)
\end{minipage}\\
\begin{minipage}[t]{\figwidth}
\centering
\includegraphics[width=\figwidth]{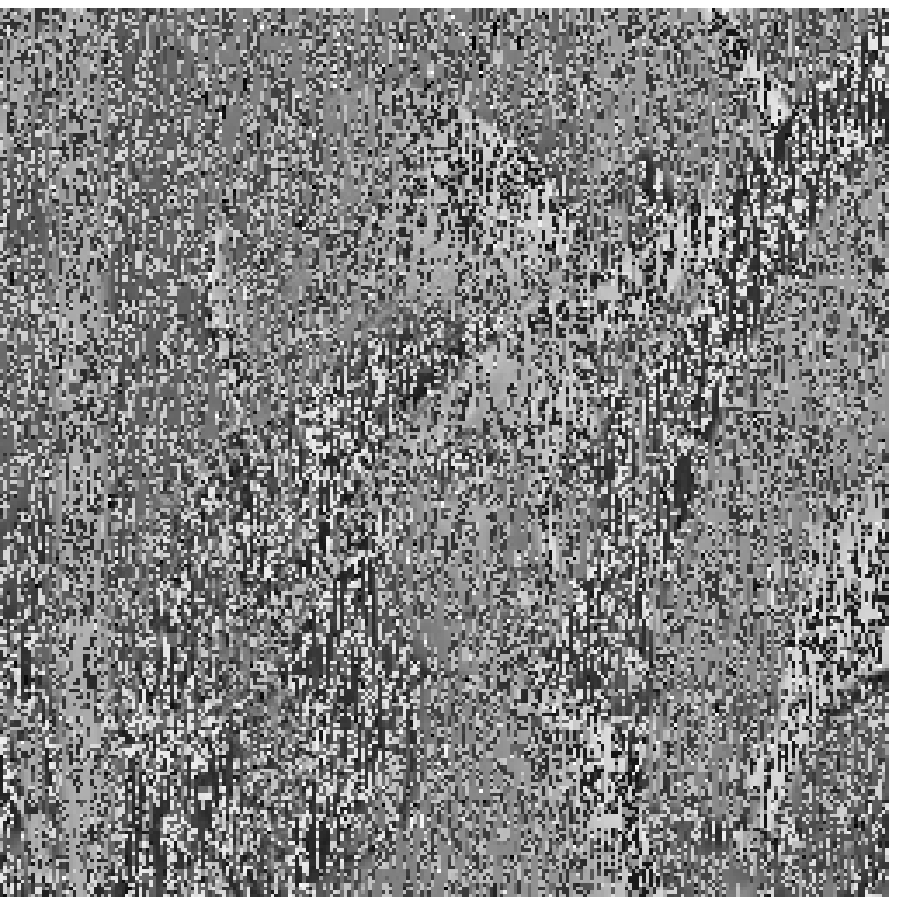}
c)
\end{minipage}
\begin{minipage}[t]{\figwidth}
\centering
\includegraphics[width=\figwidth]{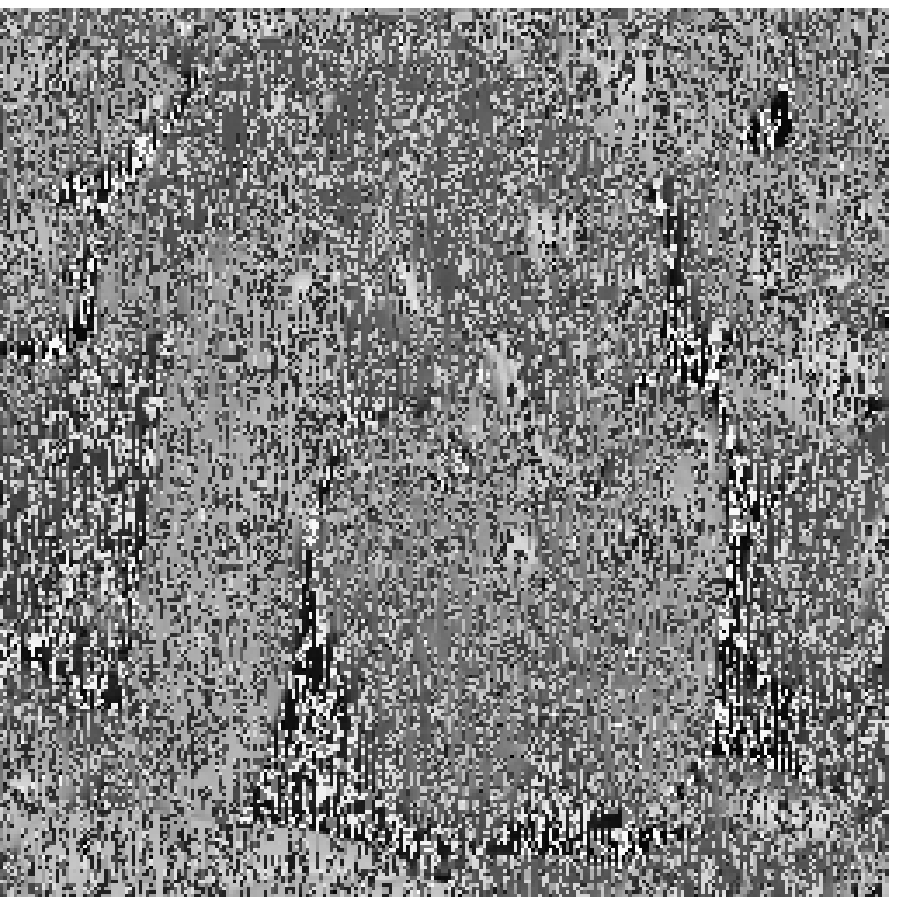}
d)
\end{minipage}
\caption{Encryption results of the DCNN-based encryption scheme on
two typical plain-images: a) the plain-image ``Lenna"; b) the
plain-image ``Peppers"; c) the cipher-image of ``Lenna"; d) the
cipher-image of ``Peppers".} \label{figure:cipherimages-DCNN}
\end{figure}

\begin{figure}[!htb]
\centering
\begin{minipage}[t]{\figwidth}
\centering
\includegraphics[width=\figwidth]{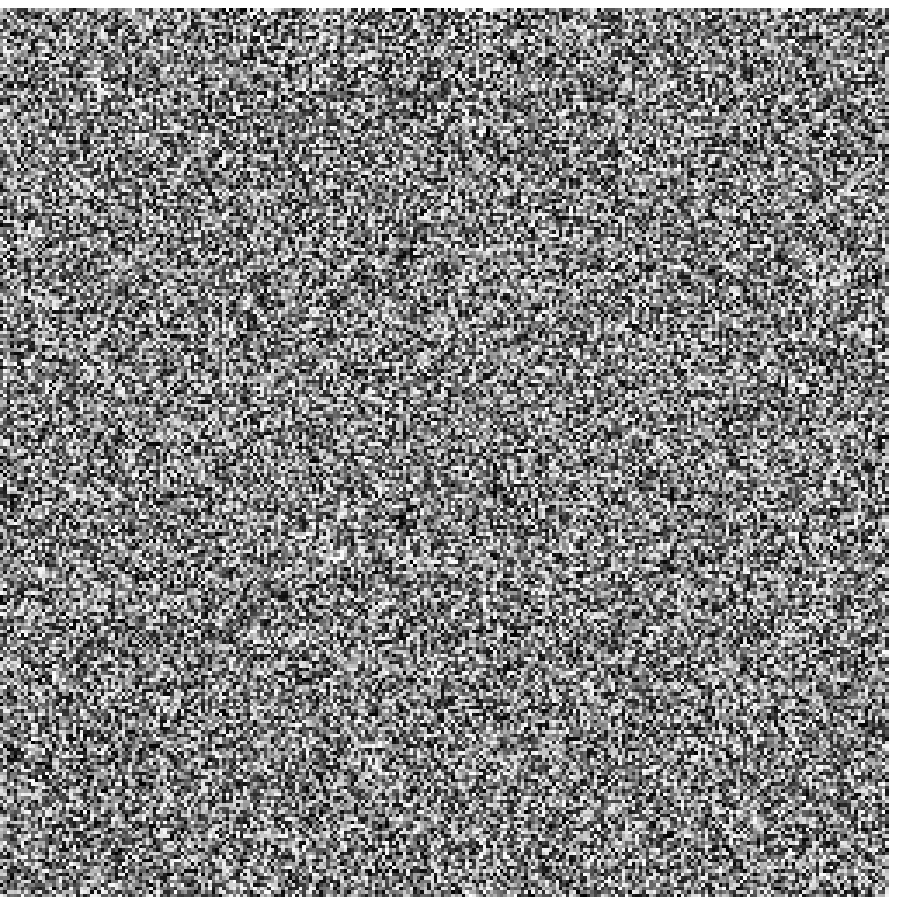}
a)
\end{minipage}
\begin{minipage}[t]{\figwidth}
\centering
\includegraphics[width=\figwidth]{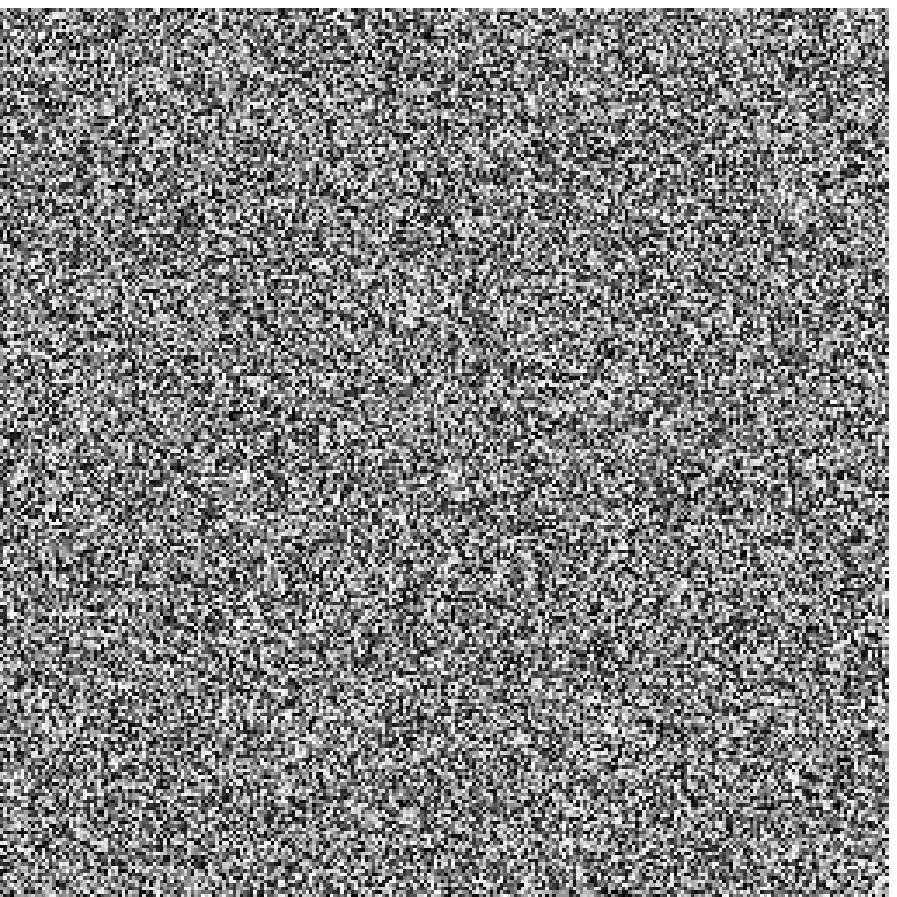}
b)
\end{minipage}
\caption{The encryption results of the encryption scheme based on
the logistic map on the two plain-images ``Lenna" and ``Peppers":
a) the cipher-image of ``Lenna"; b) the cipher-image of
``Peppers".}
\label{figure:cipherimages-logistic}
\end{figure}

\subsection{Some Inadequate Sub-Keys in the DCNN-Based Scheme}

In \cite{Wenwu:PLA2006}, it was stated that the
transfer/time-delay functions of each neuron and the numerical
algorithm itself are also part of the secret key. However, these
algorithmic details are generally embedded in the codes of the
encryption/decryption machines, so they can be reversely
engineered by analyzing the encryption/decryption machines. As a
result, they are not suitable as part of the secret key to ensure
the security of the designed cryptosystem
\cite[Sec.~1.1.7]{Schneier:AppliedCryptography96}.

Of course, if a number of candidate algorithms are embedded in the
cryptosystem, a sub-key may be introduced to secretly choose one
for encryption and decryption. With such a measure, the size of
the sub-key space is limited to the number of the candidate
algorithms, which is not large enough to make the cryptosystem
feasible in practice.

There is another problem with other sub-keys in this scheme. When
the structure of the chaotic neural network is fixed, some limits
have to be exerted on the values of the control parameters to
ensure the chaoticity of the dynamical system
\cite{HongtaoLu:ChaoticNetworks:PLA2002}. For the time step size
of the numerical algorithm, this problem also exists, as the time
step size must be small enough to approach the true dynamics of
the DCNN. This problem will reduce the key space of the encryption
scheme to some extent.

\subsection{Low Sensitivity of Encryption to Plaintexts}

As was well-known in
cryptography\cite{Schmitz:ChaoticCryptography:JFI2001}, a good
cryptosystem should be sufficiently sensitive to small changes in
the plaintext. However, this property does not hold for the two
encryption schemes proposed in
\cite{Taoxiang:PLA2006,Wenwu:PLA2006}. Observing
Eqs.~(\ref{equation:encryption1}) and
(\ref{equation:encryption2}), it is clear that one bit change in
$P_j$ will cause only one bit change in $C_j$, in the case that
the same secret key is used (i.e., both $A_j$ and $D_j$ are the
same). If there are two different bits with different values, the
distance between them modulo $L$ will also remain unchanged after
encryption.

\section{Cryptanalysis}
\label{sec:Cryptanalysis}

To facilitate the cryptanalysis given in this section, the
encryption processes of the two encryption schemes are first
unified as follows:
\begin{equation}\label{equation:encryption}
C_j=(P_j\lll_L D_j)\oplus A_j',
\end{equation}
where $A_j'=A_j$ with $L=64$ for the scheme in
\cite{Taoxiang:PLA2006} and $A_j'=(A_j\ggg_L D_j)$ with $L=32$ for
the scheme in \cite{Wenwu:PLA2006}. Apparently, if the two
sequences $\{D_j\}_{j=0}^{N-1}$ and $\{A_j'\}_{j=0}^{N-1}$ can be
reconstructed, then they can be used as an equivalent key to
decrypt the $N$ leading plain-blocks of any plaintext that is
encrypted with the same key, as follows:
\begin{equation}\label{equation:decryption}
P_j=(C_j\oplus A_j')\ggg_L D_j.
\end{equation}
This can be done by employing some special properties of circular
bit shift operations in known/chosen-plaintext attacks. Because this
idea of cryptanalysis is completely independent of the underlying
chaotic systems, it can work well for both schemes.

In the first part of this section, we give some properties of the
circular operations and then show how two very efficient and
successful attacks can be developed based on these properties.

\subsection{Some Special Properties of Circular Bit Shift
Operations}

The circular bit shift operations have the following
properties\footnote{Property~\ref{property:lll_ggg} has already
been pointed out in \cite[Sec.~7.4.1]{ShujunLi:Dissertation2003}
for another image encryption scheme based on the same
operations.}.

\begin{property}\label{property:lll_ggg}
Assume that $L,\tau\in\mathbb{Z}^+$, $x\in\mathbb{Z}$,
$a^*\in\{0,\cdots,2^\tau-1\}$ and $\tau\mid L$. If
$a=\sum_{i=0}^{L/\tau-1}\left(a^*\cdot 2^{\tau i}\right)$ and
$x\equiv 0\pmod{\tau}$, then $(a\lll_L x)=(a\ggg_L x)=a$.
\end{property}
\begin{proof}
This property is a direct consequence of the definitions of
$L$-bit left and right circular bit shift operations.
\end{proof}

\begin{property}\label{property:lll_ggg_XOR}
Assume that $L\in\mathbb{Z}^+$, $x\in\mathbb{Z}$ and
$a,b\in\{0,\cdots,2^L-1\}$. Then, $(a\lll_L x)\oplus(b\lll_L
x)=(a\oplus b)\lll_L x$ and $(a\ggg_L x)\oplus(b\ggg_L x)=(a\oplus
b)\ggg_L x$.
\end{property}
\begin{proof}
Assume that $a=\sum_{i=0}^{L-1}\left(a_i\cdot 2^i\right)$ and
$b=\sum_{i=0}^{L-1}\left(b_i\cdot 2^i\right)$. Then, $(a\lll_L
x)\oplus(b\lll_L x)=\left(\sum_{i=0}^{L-1}\left(a_i\cdot
2^{(i+x)\bmod L}\right)\right)\oplus
\left(\sum_{i=0}^{L-1}\left(b_i\cdot 2^{(i+x)\bmod
L}\right)\right)=\sum_{i=0}^{L-1}(a_i\oplus b_i)\cdot
2^{(i+x)\bmod L)}=(a\oplus b)\lll_L x$. In a similar process,
$(a\ggg_L x)\oplus(b\ggg_L x)=(a\oplus b)\ggg_L x$ can also be
proved.
\end{proof}

\begin{property}\label{property:xllla_x}
Assume that $L\in\mathbb{Z}^+\backslash\{1\}$, and
$a=\sum_{i=0}^{L-1}(a_i\cdot 2^i)\in\{0,\cdots,2^L-1\}$, where
$a_i\in\{0,1\}$. If there exists $x\in\{1,\cdots,L-1\}$ such that
$a\lll_L x=a$, then there must exist $\tau\mid\gcd(L,x)$ and
$a^*\in\{0,\cdots,2^\tau-1\}$ such that
$a=\sum_{i=0}^{L/\tau-1}(a^*\cdot 2^{\tau i})$.
\end{property}
\begin{proof}
This property is proved via mathematical induction on $x$.

When $x=1$, the condition $a\lll_L 1=a$ means the following:
$a_0=a_1$, $a_1=a_2$, $\cdots$, $a_{L-1}=a_0$, which immediately
leads to the result that $a_0=a_1=\cdots=a_{L-1}$. Then, setting
$\tau=1$ and $a^*=a_0=\cdots=a_{L-1}$, one has
$a=\sum_{i=0}^{L-1}(a^*\cdot 2^i)=\sum_{i=0}^{L/\tau-1}(b\cdot
2^{\tau i})$, where $(\tau=1)\mid\gcd(L,x)$.

Now, assume the property is true for all integers smaller than
$x\geq 2$. We will prove that it also holds for $x$. Consider two
different conditions as follows.

C1) When $x\mid L$: from the condition $a\lll_L x=a$, it follows
that $a$ can be divided into $L/x$ identical segments, each of
which has $x$ bits. Setting $\tau=x$ and
$a^*=\sum_{i=0}^{x-1}(a_i\cdot 2^i)$, we have
$a=\sum_{i=0}^{L/x-1}(a^*\cdot
2^{xi})=\sum_{i=0}^{L/\tau-1}(a^*\cdot 2^{\tau i})$, where
$(\tau=x)\mid(\gcd(L,x)=x)$.

C2) When $x\nmid L$: divide all the $L$ bits into $\lceil
L/x\rceil$ bit segments, among which the last one contains only
$\hat{x}=(L\bmod x)$ bits. That is, $a$ can be represented as
$A\cdots A\hat{A}$, where $A=a_0\cdots a_{x-1}$ and
$\hat{A}=\hat{a}_0\cdots\hat{a}_{\hat{x}-1}$. Performing $a \lll_L
x$ and comparing it with $a$ (note that $a\lll_L x=a$), one can
get $\forall i=0\sim(\hat{x}-1)$, $\hat{a}_i=a_i$. Thus, $a$
becomes $\hat{A}\check{A}\cdots \hat{A}\check{A}\hat{A}$, where
$\hat{A}=a_0\cdots a_{\hat{x}-1}$, $\check{A}=a_{\hat{x}}\cdots
a_{x-1}$ and $A=\hat{A}\check{A}$. Then, performing $a \lll_L x$
and comparing it with $a$ again, one has
$\check{A}\hat{A}=\hat{A}\check{A}=A$. This means that
$A\lll_x\hat{x}=A$. Since $\hat{x}<x$, by the assumption of the
mathematical induction, there exists $\tau\in\mathbb{Z}$ such that
$A=\sum_{i=0}^{x/\tau-1}(a^*\cdot 2^{\tau i})$, where
$\tau\mid\gcd(x,\hat{x})$ and $a^*\in\{0,\cdots,2^\tau-1\}$. Since
$\tau\mid\hat{x}$ also holds, one has
$\hat{A}=\sum_{i=0}^{\hat{x}/\tau-1}(a^*\cdot 2^{\tau i})$.
Finally, therefore, $a=\sum_{i=0}^{L/\tau-1}(a^*\cdot 2^{\tau
i})$.

The above induction completes the proof of the property.
\end{proof}
\begin{remark}
In Property~\ref{property:xllla_x}, if one changes $a\lll_L x=a$
to another form, $a=a\lll_L(L-x)$, the condition of $\tau$ will
become $\tau\mid\gcd(L,L-x)$. This is actually equivalent to
$\tau\mid\gcd(L,x)$, as $\gcd(L,x)=\gcd(L,L-x)$.
\end{remark}

Combining Properties~\ref{property:lll_ggg} and
\ref{property:xllla_x}, one can easily derive the following
theorem.

\begin{theorem}\label{theorem:xllla_y}
Assume that $L\in\mathbb{Z}^+\backslash\{1\}$, $x\in\mathbb{Z}$
and $a,b\in\{0,\cdots,2^L-1\}$. The equation $(a\lll_L x)=b$ ($x$
is the unknown) has more than one solution modulo $L$ if and only
if there exists $\tau<L$, $\tau\mid L$ and
$a^*\in\{0,\cdots,2^\tau-1\}$ such that
$a=\sum_{i=0}^{L/\tau-1}(a^*\cdot 2^{\tau i})$.
\end{theorem}
\begin{proof}
The ``if" and ``only if" parts of this theorem are direct
consequences of Properties~\ref{property:lll_ggg} and
\ref{property:xllla_x}, respectively.
\end{proof}

An alternative form of Theorem~\ref{theorem:xllla_y} is as
follows.

\begin{theorem}\label{theorem:xllla_y_one}
Assume that $L\in\mathbb{Z}^+\backslash\{1\}$, $x\in\mathbb{Z}$
and $a,b\in\{0,\cdots,2^L-1\}$. The equation $(a\lll_L x)=b$ ($x$
is the unknown) has only one solution modulo $L$ if and only if
there does not exist $\tau<L$, $\tau\mid L$ and
$a^*\in\{0,\cdots,2^\tau-1\}$ satisfying
$a=\sum_{i=0}^{L/\tau-1}(a^*\cdot 2^{\tau i})$.
\end{theorem}

When $\tau<L$, $a=\sum_{i=0}^{L/\tau-1}(a^*\cdot 2^{\tau i})$
actually means that $a$ can be represented by repeated bit
patterns. For example, when $L=8$, $\tau=4$ and $a^*=(1001)_2=9$,
one has $a=(10011001)_2=153$, where $(\cdots)_2$ denotes the
binary representation (the same below).

\subsection{Chosen-Plaintext Attack}

In this attack, two plaintexts can be deliberately chosen to
ensure that all elements in $\{D_j\}$ and $\{A_j'\}$ are uniquely
determined. By choosing a plaintext such that $P_j^{(1)}=0$ or
$2^{L}-1$, $\forall j=0\sim(N-1)$, one obtains
$(P_j^{(1)}\lll_{L}D_j)=P_j^{(1)}$ and further gets
$A_j'=P_j^{(1)}\oplus C_j^{(1)}$. After recovering
$\{A_j'\}_{j=0}^{N-1}$, one may choose another plaintext such that
each $P_j^{(2)}$ cannot be represented by repeated bit patterns,
for example, $P_j^{(2)}=152=(10011000)_2$ when $L=8$. Then, by
Theorem~\ref{theorem:xllla_y_one}, the value of $D_j$ can always
be uniquely determined by solving the following equation:
\[
(P_j^{(2)}\lll_L D_j)=C_j^{(2)}\oplus A_j'.
\]

\subsection{Differential Known-Plaintext Attack}

When the same key is used to encrypt two plaintexts,
$\{P_j^{(1)}\}_{j=0}^{N-1}$ and $\{P_j^{(2)}\}_{j=0}^{N-1}$, using
Eq.~(\ref{equation:encryption}) and
Property~\ref{property:lll_ggg_XOR} one can easily deduce the
following equality:
\begin{eqnarray}
C_j^{(1)}\oplus C_j^{(2)} & = & \left(P_j^{(1)}\lll_{L}
D_j\right)\oplus\left(P_j^{(2)}\lll_{L} D_j\right)\nonumber\\
& = & \left(P_j^{(1)}\oplus P_j^{(2)}\right)\lll_{L}
D_j.\label{equation:XORdiff}
\end{eqnarray}
The above equation means that $\{A_j'\}_{j=0}^{N-1}$ are
completely circumvented in a differential attack. Then, one can
try to determine the value of $D_j$ by searching all $L$ possible
values. From Theorem~\ref{theorem:xllla_y_one}, the value $D_j$
can be uniquely determined if $P_j^{(1)}\oplus P_j^{(2)}$ cannot
be represented in repeated bit patterns. After obtaining $D_j$,
one can further get the value of $A_j'$ as follows:
\begin{equation}
A_j'=\left(P_j^{(1)}\lll_{L} D_j\right)\oplus C_j^{(1)}.
\end{equation}

Now, let us find the probability that the value of each $D_j$
cannot be uniquely determined by solving
Eq.~(\ref{equation:XORdiff}), i.e., the probability that
$P_j^{(1)}\oplus P_j^{(2)}$ can be represented as repeated bit
patterns. Under the assumptions that $P_j^{(1)}\oplus P_j^{(2)}$
has a uniform distribution over $\{0,\cdots,2^{L}-1\}$ and that
any two differential values are independent of each other, this
probability can be calculated to be
\begin{equation}
p=\frac{\sum_{\tau<L,\;\tau\mid L}2^\tau}{2^{L}}.
\end{equation}
Then, it can be easily calculated that $p\approx 2^{-16}$ when
$L=32$ and $p\approx 2^{-32}$ when $L=64$. In practice, this
probability is generally larger than the theoretical value due to
the non-uniform distribution of the plaintext and the correlation
existing in two differential plaintexts. To the advantage of the
attacker, our experiments have shown that this probability is
still very small in most cases. The small probability ensures that
it is a high-probability event to uniquely determine the value of
$D_j$ with only two known plaintexts and their corresponding
ciphertexts.

To evaluate the performance of the differential plaintext attacks,
some experiments were carried out when a number of natural images
are chosen as plaintexts. Consider the case of the 2-neuron DCNN
shown in Eq.~(\ref{eq:twoneurons}) with the same configurations
set in Sec.~\ref{ssec:nonuniformdistribution}. With the two
plain-images and the corresponding cipher-images shown in
Fig.~\ref{figure:cipherimages-DCNN}, we reconstructed
$\{D_j\}_{j=0}^{16383}$ and $\{A_j'\}_{j=0}^{16383}$. In all the
$16384$ elements of each sequence, only two 1's could not be
uniquely determined, which is about $0.012\%$ ($\approx 2^{-13}$).
Then, the two reconstructed sequences $\{D_j\}_{j=0}^{16383}$ and
$\{A_j'\}_{j=0}^{16383}$ were used to decrypt a cipher-image shown
in Fig.~\ref{figure:breakcipherimage-DCNN}a (which corresponds to
a plain-image ``House"). The recovered plain-image is given in
Fig.~\ref{figure:breakcipherimage-DCNN}b. One can see that the
breaking performance is nearly perfect.

\begin{figure}[!htb]
\centering
\begin{minipage}[t]{\figwidth}
\centering
\includegraphics[width=\figwidth]{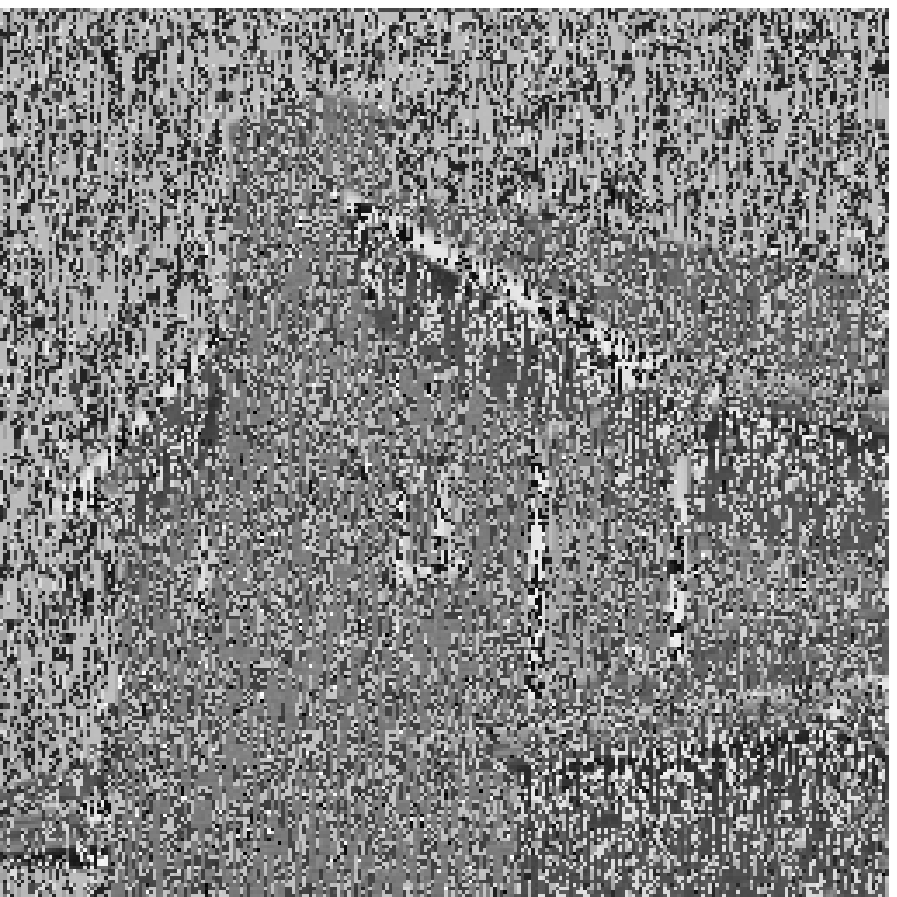}
a)
\end{minipage}
\begin{minipage}[t]{\figwidth}
\centering
\includegraphics[width=\figwidth]{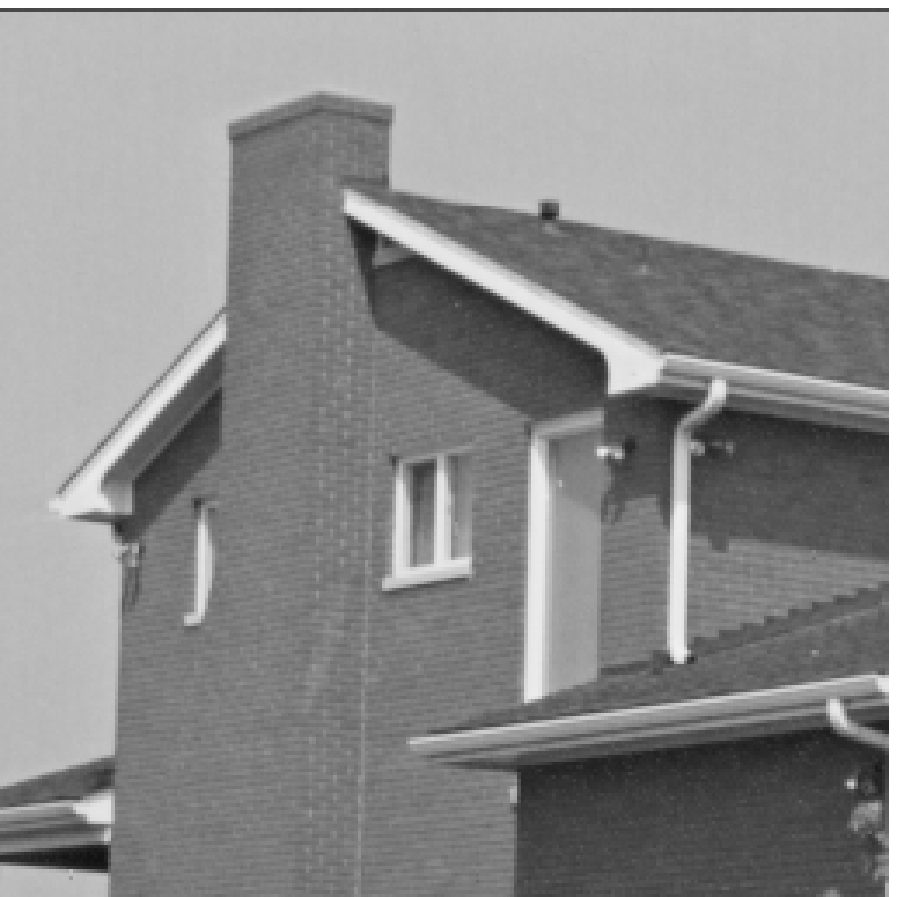}
b)
\end{minipage}
\caption{A near-perfect breaking result of the differential
known-plaintext attack on the DCNN-based scheme: a) the
cipher-image corresponding to a plain-image ``House"; b) the
decrypted plain-image.} \label{figure:breakcipherimage-DCNN}
\end{figure}

The same experiments were also carried out on the scheme based on
the logistic map with the same known plain-images and the
corresponding cipher-images shown in
Fig.~\ref{figure:cipherimages-logistic}. As analyzed above, in
this case the probability that each value of $D_j$ and $A_j'$
cannot be uniquely determined is estimated to be $2^{-32}$.
Considering there are only $256\times 256/8=2^{13}$ plain-blocks,
one can expect that all elements in $\{D_j\}_{j=0}^{8191}$ and
$\{A_j'\}_{j=0}^{8191}$ will be uniquely determined in very high
probability, thus leading to a perfect breaking of the
plain-image. Our experiments well agreed with this expectation.
Figure~\ref{figure:breakcipherimage-logistic} shows the breaking
result on the plain-image ``House".

\begin{figure}[!htb]
\centering
\begin{minipage}[t]{\figwidth}
\centering
\includegraphics[width=\figwidth]{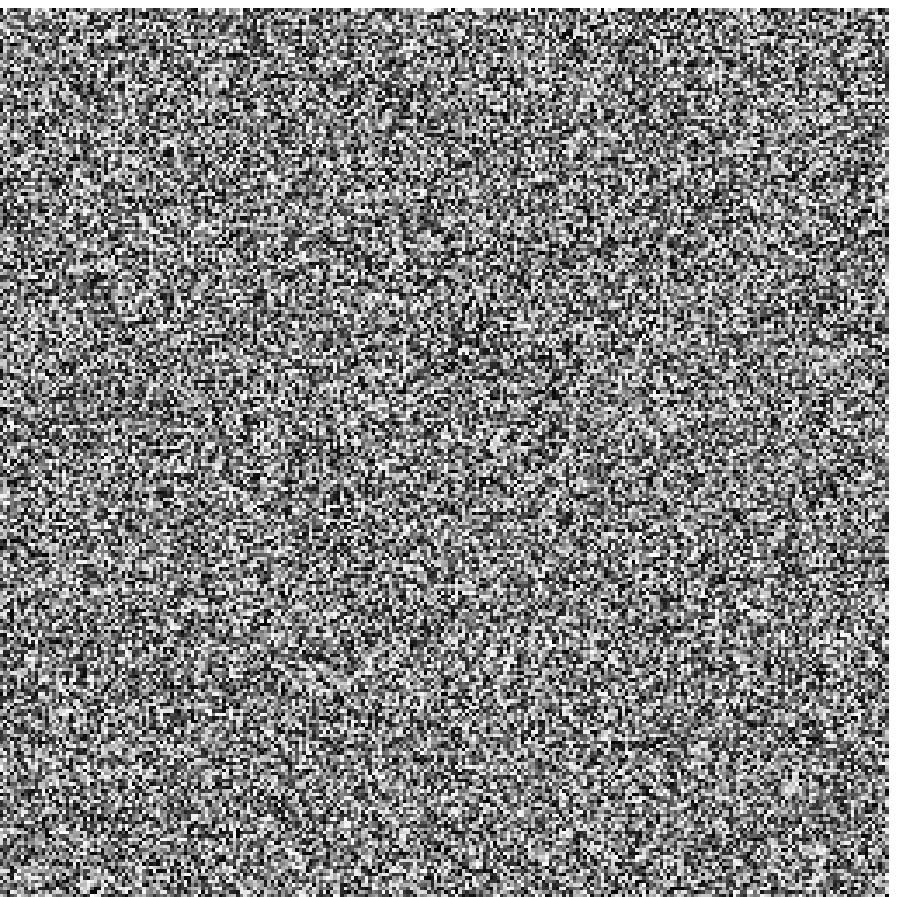}
a)
\end{minipage}
\begin{minipage}[t]{\figwidth}
\centering
\includegraphics[width=\figwidth]{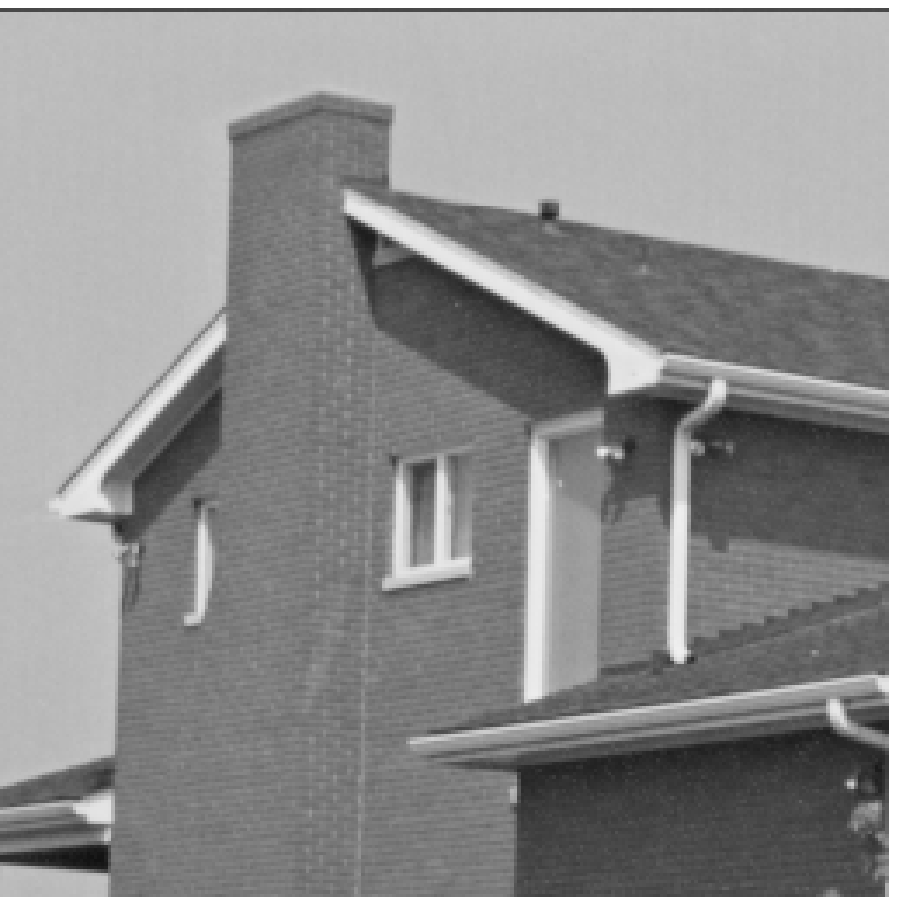}
b)
\end{minipage}
\caption{A perfect breaking result of the differential
known-plaintext attack on the encryption scheme based on the
logistic map: a) the cipher-image corresponding to ``House"; b)
the recovered plain-image.}
\label{figure:breakcipherimage-logistic}
\end{figure}

\section{Conclusions}

This paper has analyzed the security of two chaotic encryption
schemes based on circular bit shift and XOR operations. It has
been found that these two schemes are insecure against the
differential known-plaintext attack and the chosen-plaintext
attack, in which only two known/chosen plaintexts are required to
achieve a perfect breaking performance. Moreover, some other
security problems existing in the two encryption schemes have been
pointed out. Our cryptanalytic results suggest that the two
encryption schemes should be further enhanced before they can be
used in applications requiring a high level of security.

\begin{ack}
This research was partially supported by The Hong Kong Polytechnic
University's Postdoctoral Fellowships Program under grant no.
G-YX63 and by Ministerio de Ciencia y Tecnologia of Spain under
research grant SEG2004-02418. The authors would like to thank
Prof. L.~F. Shampine at the Southern Methodist University (USA)
for his help in solving for numerical solutions to the delay
differential equations of the DCNN.
\end{ack}

\bibliographystyle{elsart-num}
\bibliography{DCNN}

\end{document}